\documentclass[aps, prresearch, reprint, nofootinbib, longbibliography]{revtex4-2}

\usepackage[utf8]{inputenc}
\usepackage[T1]{fontenc} 
\usepackage[T1]{fontenc}
\usepackage[utf8]{inputenc}
\usepackage[dvipsnames]{xcolor} 

\definecolor{myblue}{rgb}{0.06,0.30,0.55}
\definecolor{myblue1}{rgb}{0.06,0.60,0.55}
\definecolor{mydeepblue}{RGB}{0,0,139}
\definecolor{Dgreen}{RGB}{0, 100, 0}
\definecolor{blue2}{RGB}{10, 52, 96}
\usepackage{siunitx}
\usepackage{upgreek}  
\usepackage{amsmath}
\usepackage{amssymb}
\usepackage{amstext}
\usepackage{amsbsy}
\usepackage{mathtools}
\usepackage{mathrsfs}
\usepackage{physics} 
\usepackage{dsfont}
\usepackage{bbm}
\usepackage{bm}
\usepackage{amsthm}
\usepackage{extarrows}
\usepackage[retainorgcmds]{IEEEtrantools} 
\usepackage{siunitx} 
\usepackage[T1]{fontenc}
\usepackage[utf8]{inputenc}
\usepackage{graphicx} 
\usepackage{dcolumn}
\usepackage{booktabs}
\usepackage{multirow}
\usepackage{diagbox}
\usepackage{float}
\usepackage{subfigure} 
\usepackage{textcomp}
\usepackage{mdwlist}
\usepackage{lipsum}
\usepackage{textcomp}
\usepackage{url}
\usepackage[title]{appendix}
\usepackage{etoolbox}

\BeforeBeginEnvironment{appendices}{%
	\renewcommand{\thesection}{\Alph{section}}%
	\renewcommand{\section}{%
		\secdef{\@Appendix}{\@sAppendix}%
	}%
	\newcommand{\@Appendix}[2][?]{%
		\refstepcounter{section}%
		\vspace{0.5em}
		{\centering\large\bfseries Appendix \thesection: #2\par}%
		\addcontentsline{toc}{section}{Appendix \thesection}%
		\markright{Appendix \thesection}%
	}%
}

\usepackage{hyperref}
\hypersetup{%
	plainpages=true,
	breaklinks=true,
	hypertexnames=false,
	pageanchor=true,
	colorlinks=true,
	linkcolor={blue},
	citecolor={blue}, 
	urlcolor={blue},  
	anchorcolor={black}
}

\begin{document}

			
			
			
			\title{Squeezing-enhanced dual-channel interference for ground-state cooling of a levitated micromagnet with low quality factor}

	\author{Lei Chen$^{1,2}$}
	\author{Zhe-qi Yang$^{1,2}$}
	\author{Liang Bin$^{1,2}$}
	\author{Zhi-Rong Zhong$^{1,2}$}
	\email{zhirz@fzu.edu.cn}
	\date{\today }
	\address{$^1$Fujian Key Laboratory of Quantum Information and Quantum Optics, Fuzhou University, Fuzhou 350108, China}
    \address{$^2$College of Physics and Information Engineering, Fuzhou University, Fuzhou 350108, China}

\begin{abstract}
	Cooling the center-of-mass (CM) motion of a macroscopic oscillator to its quantum ground state is a fundamental prerequisite for testing quantum mechanics at macroscopic scales.  However, achieving this goal is currently hindered by the stringent requirement for an ultrahigh mechanical quality factor ($Q_c$). Here, we propose a dual-channel cooling scheme based on squeezing-enhanced quantum interference within a hybrid levitated cavity-magnomechanical system to overcome this limitation.
	By synergizing squeezing effects with quantum interference between the magnon-CM and cavity-CM channels, our scheme simultaneously suppresses Stokes (heating) scattering while enhancing anti-Stokes (cooling) scattering.~We demonstrate that this cooling mechanism reduces the critical $Q_c$ required for ground-state cooling by three orders of magnitude, making it achievable in the experimentally accessible regime of $Q_c \sim 10^4$. Furthermore, the net cooling rate is enhanced by nearly 180-fold compared to that of conventional single-channel cooling. This improvement is accompanied by a two orders of magnitude reduction in both the steady-state CM occupancy and the cooling time. Importantly, this enhanced performance remains robust even deep within the unresolved-sideband regime. Our results provide a feasible path toward preparing macroscopic quantum states by actively controlling the cooling dynamics, thereby relaxing the constraints on intrinsic material properties.
\end{abstract}

\keywords{cavity magnomechanics, quantum cooling, optical parametric amplification, quantum interference} 
\maketitle

\section{Introduction}

Cooling the center-of-mass (CM) motion of a macroscopic object to the quantum ground state is a crucial prerequisite for testing quantum mechanics at macroscopic scales~\cite{sciadv1603150,Cooling367,dark}, searching for new physics such as dark matter~\cite{PhysRevD.110.115029,PhysRevLett.125.181102}, and enabling ultra-high-precision metrology beyond the standard quantum limit~\cite{PhysRevApplied.8.034002}.
However, achieving this goal for objects at the micro- to millimeter scale remains a challenge. While levitation techniques such as optical~\cite{Seberson2020Sympathetic}, magnetic~\cite{PhysRevLett.119.167202}, and Paul~\cite{PhysRevB.101.134415} traps have enabled effective cooling and trapping of nanoparticles or particles of nanogram masses, scaling to larger masses presents a formidable obstacle.

Recent progress in magnetic levitation~\cite{PhysRevLett.124.093602,PhysRevApplied.11.044041,PhysRevB.96.134419, PhysRevA.96.063810,levitation} provides a promising platform for stably suspending magnetic particles ranging from micrometers to millimeters, paving the way for cooling their CM motion~\cite{Pontin2023simultaneous,prr,PhysRevLett128013602}. However, conventional cooling methods~\cite{PhysRevLett128013602,PhysRevLett.126.193602,Nc580,PhysRevLett.122.123601,PhysRevLett.110.153606,PhysRevLett.117.173602}, such as feedback cooling~\cite{PhysRevResearch.6.033345,PhysRevA.103.L051701,Nc373,Nc595}, sideband cooling~\cite{ PhysRevLett.123.153601}, passive cooling~\cite{Nature7289}, and continuous measurement cooling---generally require a high mechanical quality factor (\( Q_c \)) of the CM mode. Theoretical proposals often demand extremely high \( Q_c \) values, on the order of \( 10^{8} \)~\cite{PhysRevA.108.023523} to \( 10^{11} \)~\cite{PhysRevLett128013602}, to ensure the energy coherence time surpasses the cooling timescale.
These requirements necessitate exceptionally stringent conditions, such as ultra-high vacuum with $P$$\sim$1$\times 10^{-10}$\,\si{\milli\bar}~\cite{12, 2013Thermal}, posing significant experimental challenges. 
In stark contrast, experimentally measured \( Q_c \)
values of micron-scale micromagnet spheres in magnetic levitation typically range from only $10^3$~\cite{PhysRevA.108.063511} to $10^7$~\cite{PhysRevLett.124.163604,PhysRevLett.131.043603}. 
This stark disparity between theoretical requirements and experimental capabilities underscores the critical limitation of existing cooling methods: their passive reliance on intrinsically extreme physical conditions. Moreover, realistic experimental systems are inevitably subject to additional noise sources such as eddy-current losses~\cite{PhysRevA.108.063511} and magnetic field fluctuations~\cite{PhysRevA.108.023523}, which introduce considerable thermal noise and further limit cooling efficiency.

Given this challenge, the development of novel cooling mechanisms that remain efficient under low $Q_c$ conditions is paramount.~To date, various quantum techniques---such as intracavity squeezing~\cite{PhysRevLett.124.103602, Gan2019}, injected squeezed light~\cite{Asjad2019,PhysRevA.94.051801,Nat2017,Aasi2013}, optical parametric amplification (OPA)~\cite{qipOPA2023}, cavity-assisted cooling and quantum interference~\cite{PhysRevA.87.025804}---have been proposed to enhance cooling performance. However, these schemes fail to address the challenges of cooling CM modes in levitated systems under low $Q_c$ conditions.
Crucially, protocols that exploit the unique degrees of freedom in hybrid levitated systems to actively circumvent the low-$Q_\text{c}$ limitation remain scarce.
Therefore, innovative protocols that can effectively enhance CM cooling and significantly accelerate the cooling rate in levitated systems are urgently needed.


To address this challenge, we propose a dual-channel cooling scheme based on squeezing-enhanced quantum interference within a hybrid levitated cavity-magnomechanical (CMM) system.\hspace{0.5em}The
proposed mechanism, named the Cavity-CM (CCM) and Magnon-CM (MCM)  Interference (CMI) dual-channel cooling mechanism. 
By exploiting squeezing-enhanced synergistic quantum interference between the CCM and MCM channels, our scheme simultaneously suppresses Stokes (heating) scattering and enhances anti-Stokes (cooling) scattering.
A key breakthrough of this work is that the CMI cooling mechanism \textbf{\textit{reduces}}
the required $Q_c$ by  \textbf{\textit{three orders of magnitude}}, enabling efficient ground-state cooling in an experimentally accessible regime.
Moreover, the enhanced cooling performance remains robust even deep in the unresolved-sideband regime, thereby overcoming a key limitation of conventional sideband cooling.

Our numerical simulations demonstrate the effectiveness of the CMI dual-channel cooling mechanism. 
Compared to the conventional MCM single-channel cooling mechanism, the CMI dual-channel cooling mechanism yields a nearly \textbf{\textit{180-fold enhancement}} in the net cooling rate---a dramatic improvement attributable to quantum interference effects.
It also reduces the steady-state CM occupancy and the cooling time by two orders of magnitude. 
Notably, the stringent requirement on $Q_c$ is relaxed, allowing efficient ground-state cooling for systems with
$Q_c\sim 10^4$.
Our work thus establishes an experimentally feasible route for observing macroscopic quantum phenomena by actively controlling cooling dynamics rather than passively relying on intrinsic material properties.
\begin{figure*}[t]  
	\centering	 
	\includegraphics[width=7in]{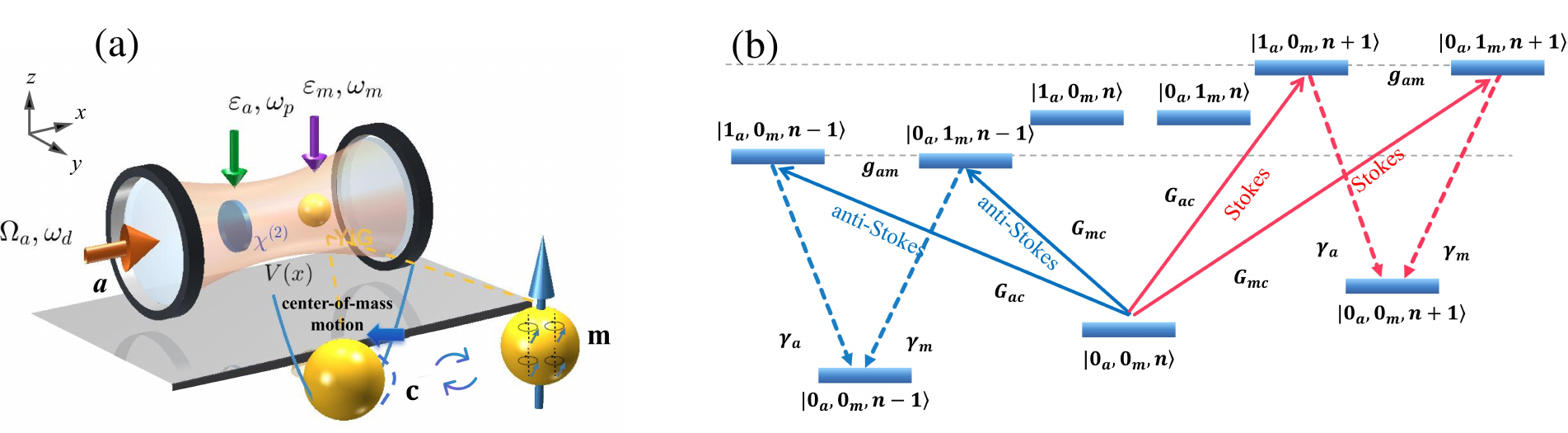}		
		\caption{(a) Sketch of the hybrid CMM system: A YIG sphere is levitated and trapped at the potential minimum $V(x)$ in a cavity. The cavity contains a YIG sphere and a second-order nonlinear crystalline medium. 
		(b) Schematic illustration of the quantum interference between scattering channels in the CMI cooling mechanism.} 
	\label{fig:1}
\end{figure*}

The paper is organized as follows.
In Sec.~\ref{II}, we introduce the theoretical model of the hybrid levitated CMM system. Section~\ref{III} is devoted to deriving the quantum Langevin equations.~The steady-state solutions and the linearized Hamiltonian are presented in Sec.~\ref{4}. In Sec.~\ref{V}, we analyze the quantum noise spectra and net cooling rates for the two cooling mechanisms. Sec.~\ref{VI} provides a comprehensive evaluation of the cooling performance through numerical simulations of the spectral density, net cooling rate, and steady-state CM occupancy. We discuss the experimental feasibility of our proposal in Sec.~\ref{VII}. Finally, Sec.~\ref{VIII} summarizes our main results.\\\\

\section{Theoretical model} \label{II}

We consider a hybrid levitated CMM system~\cite{Nonreciprocity,PhysRevA.110.013710}, which consists of a cavity containing a levitated YIG sphere. The YIG sphere is chosen for its exceptional properties, such as high spin density and low magnetic loss~\cite{RevModPhys.85.623,Zhang:22,2016Cavity}. It has a diameter $d > 20~\mathrm{\upmu m}$ and is confined in a low-frequency trap ($\sim \SI{50}{\kilo\hertz}$).
A uniform bias magnetic field is applied to excite the magnon mode within the sphere, while the CM motion of the sphere is confined by a harmonic potential.
We focus on the Kittel mode---the lowest-order, spatially uniform magnon mode governing the collective spin precession in the sphere~\cite{2016Cavity}.
This single-mode approximation is justified when the diameter $d$ of the YIG sphere is much smaller than the wavelength of the cavity field $\lambda_a$. Under this condition ($d \ll \lambda_a$), the cavity magnetic field is nearly uniform across the sphere, leading to a vanishing spatial overlap with higher-order magnon modes and thus negligible interactions between the cavity mode and these modes~\cite{PhysRevLett.123.107702,PhysRevB.94.224410}.
The Kittel mode can coherently couple to multiple physical subsystems, including the mechanical vibrational (phonon) mode of the sphere, the photon mode, and the CM motional mode, as illustrated in Fig.~\ref{fig:1}(a).
Under this condition, the total Hamiltonian of the system is given by
\begin{align}
\hat{H} &= \hat{H}_0 + \hat{H}_{\text{int}} + \hat{H}_d. \label{eq:total_H}
\end{align}
The free Hamiltonian $\hat{H}_0$ in Eq.~\eqref{eq:total_H} is 
\begin{align}
\hat{H}_0/\hbar &= \omega_a \hat{a}^\dagger \hat{a} + \omega_m \hat{m}^\dagger \hat{m} + \frac{\omega_c}{2} \left( \hat{x}^2 + \hat{p}_x^2 \right), \label{eq:H0}
\end{align}
where $\hat{a}$ ($\hat{a}^\dagger$) and $\hat{m}$ ($\hat{m}^\dagger$) denote the annihilation (creation) operators for the cavity and magnon modes, with frequencies $\omega_a$ and $\omega_m$, respectively.
The magnon frequency $\omega_m = \gamma B_0$ can be tuned via an external magnetic field $B_0$, where $\gamma/2\pi= 28 \, \mathrm{GHz/T}$ is the gyromagnetic ratio of the YIG sphere.
Note that the frequency of the Kittel magnon mode can be precisely tuned to near resonance with the cavity mode ($\omega_a \approx \omega_m$) by adjusting the external magnetic field.
The CM motion of the YIG sphere is described by the displacement and momentum operators $\hat{x}$ and $\hat{p}_x$, oscillating harmonically with the trap frequency $\omega_c$~\cite{PhysRevA.110.013710,prr, PhysRevLett128013602}. The displacement
operator is quantized as:
$	\hat{x} = x_{\scriptscriptstyle\text{ZPM}}(\hat{c} + \hat{c}^\dagger), $
where $\hat{c}$ ($\hat{c}^\dagger$) is the annihilation (creation) operator for the CM modes, 
Here, $x_{\scriptscriptstyle\text{ZPM}} = \sqrt{\frac{\hbar}{2{\rho_m}V\omega_{c}}}$ is zero-point fluctuations of the CM mode in the trapping potential,
$\rho_m = 5170\ \text{kg}\ \cdot\text{m}^{-3}$ is the mass density of the YIG sphere, $V$ is the volume of the sphere.
The magnon--photon coupling is mediated by the magnetic dipole interaction
$
\hat{H}_{\text{int}} = -\vec{\hat{m}} \cdot \vec{B}_{\text{cav}}. \label{eq:magnetic_dipole}
$
The cavity mode is polarized along the $\hat{y}$-direction, with a spatially modulated magnetic field
$
\vec{B}_{\text{cav}}(x) = \hat{y} B_0 \cos(kx), \label{eq:B_cav}
$
where $k$ is the wave number of the microwave field propagating along the $\hat{x}$-axis~\cite{PhysRevLett128013602}, and the bias magnetic field along the \textit{z}-axis.

The Hamiltonian $\hat{H}_{\text{int}}$ in Eq.~\eqref{eq:total_H}, which governs the interaction between the magnon mode and the photon mode, is given by
\begin{align}
\hat{H}_{\text{int}}/\hbar &= g_{am} \left( \hat{a}^\dagger \hat{m} + \hat{a} \hat{m}^\dagger \right) \cos(kx), \label{eq:H_int}
\end{align}
where the magnon--photon coupling strength~\cite{PhysRevLett128013602, PhysRevLett.113.156401} is given by $
g_{am} = \frac{\gamma}{2} \sqrt{\frac{\hbar \omega_a \mu_0}{V_a}} \sqrt{2 \rho_s V s}. \label{eq:g_am}
$
Here, $\mu_0$ is the vacuum permeability, $V_a$ is the volume of the cavity mode,  $\rho_s = \SI{4.22e27}{\per\cubic\meter}$  is the spin density, and $s = 5/2$ is the ground-state spin quantum number. The coupling rate $g_{am}$ is experimentally tuned by varying the position of the YIG sphere within the cavity mode via the external magnetic field $B_0$~\cite{PhysRevLett.113.156401}.

Furthermore, we introduce three external driving fields, the total driving Hamiltonian $\hat{H}_d$ in Eq.~\eqref{eq:total_H} thus includes three distinct components:
\begin{align}
	\hat{H}_{d}/\hbar = &i\Omega_{a}\left(\hat{a}^{\dagger} e^{-i\omega_{d} t}-\hat{a} e^{i\omega_{d} t}\right)  + \varepsilon_{a}\hat{a}^{\dagger 2} e^{-i\omega_{p} t} \nonumber \\
	& + {\varepsilon^*_{a}}\hat{a}^{2} e^{i\omega_{p} t} + i\varepsilon_{m}\left(\hat{m}^{\dagger} e^{-i\omega_{0} t}-\hat{m} e^{i\omega_{0} t}\right),
	\label{eq:H_drive}
\end{align}
where $|\Omega_a| = \sqrt{2 \gamma_a P_d/(\hbar \omega_d)}$ is the drive amplitude of the cavity  field with frequency $\omega_d$, $\gamma_a$ is the decay rate of the cavity, and $P_d$ is the input power.
The second term in Eq.~\eqref{eq:H_drive} corresponds to an intracavity squeezing term, which is realized by a pump field with frequency $\omega_p=2\omega_d$ and amplitude $\varepsilon_a$.
This field drives the nonlinear medium inside the cavity, inducing an OPA process that generates photon pairs. 
Here, the amplitude $\varepsilon_a$ serves as the effective squeezing parameter, which we express in the complex form as $\varepsilon_a = i\Lambda e^{i\theta}/2$, where $\Lambda$ and $\theta$ are the amplitude and phase of the intracavity squeezing, respectively.
The third term in Eq.~\eqref{eq:H_drive} describes the direct driving of the magnon mode, with amplitude
$
\varepsilon_m = \gamma \sqrt{\frac{5N }{4}}B_0,
\label{eq:drive_strength}
$
where $B_0$ is the field amplitude with frequency $\omega_0$, and $N = \rho V$ represents the total number of spins.

Under the resonance condition ($\omega_d = \omega_0$) and expanding $\cos(kx)$ to first order around the potential minimum $x_0$, i.e., $\cos(k\hat{x})\rightarrow\cos(kx_0)-k\sin(kx_0)(\hat{x}-x_0)$. The total Hamiltonian in the rotating frame can be written as\begin{align}
\hat{H}/\hbar
&= \Delta_a \hat{a}^\dagger \hat{a}
+ \Delta_m \hat{m}^\dagger \hat{m}
+ \omega_c \hat{c}^\dagger \hat{c} \nonumber \\
&\quad+ \bigl( \hat{a}\hat{m}^\dagger + \hat{a}^\dagger \hat{m} \bigr)
\bigl[ g_{am} \cos(kx_0) - g_\text{amc} \sin(kx_0)
\bigl( \hat{c}^\dagger + \hat{c} \bigr) \bigr] \nonumber \\
&\quad+ \left( {\varepsilon_{a}}\hat{a}^{\dagger 2} + {\varepsilon^*_{a}}\hat{a}^2  \right)
+ i\Omega_a \left( \hat{a}^{\dagger} - \hat{a} \right)
+ i\varepsilon_m \left( \hat{m}^{\dagger} - \hat{m} \right),
\label{eq:H_rotating}
\end{align}
where the detunings of the photon and magnon excitations relative to the driving field are $\Delta_j = \omega_j - \omega_d$ ($j = a, m$). The magnon--photon--CM coupling strength is
$g_{\mathrm{amc}} = g_{am} k x_{\scriptscriptstyle{\mathrm{ZPM}}}
= \frac{1}{2} \hbar k \gamma \sqrt{\frac{\omega_a \mu_0 \rho_s s}{\rho_m \omega_c V_a}}
\label{eq:g_amc}
$~\cite{PhysRevLett.113.156401}.
Notably, the coupling strength $g_{\text{amc}}$ is independent of the volume of the magnetic sphere, which is a key advantage for scaling to larger masses.

\section{Quantum langevin equations} \label{III}

To describe the dissipative dynamics of our system, we begin with the Lindblad master equation:
\begin{equation}
\dot{\rho} = -\frac{i}{\hbar}[\hat{H}, \rho]
- \sum_{\hat{o} = \hat{a}, \hat{m}, \hat{c}} \frac{\gamma_o}{2}
\Bigl[ (\bar{n}_o + 1) \mathcal{L}(\hat{o}) \rho
+ \bar{n}_o \mathcal{L}(\hat{o}^\dagger) \rho \Bigr],
\label{eq:master_eq}
\end{equation}
where the Lindblad superoperator is \( \mathcal{L}(\hat{o})\rho = (\hat{o}^\dagger \hat{o}\rho + \rho \hat{o}^\dagger \hat{o} - 2\hat{o}\rho \hat{o}^\dagger)/2 \). Here, \( \hat{o} \) (\( \hat{o} \in \{\hat{a}, \hat{m}, \hat{c}\} \)) denotes the annihilation operator for the relevant bosonic modes, \( \rho \) is the density operator of the system, and \( \bar{n}_o = 1/[\exp(\hbar\omega_o/k_B T)-1] \) is the mean thermal occupation number, with \( k_B \) being the Boltzmann constant, \( T \) is the environmental temperature, and \( \gamma_o \) and \( \omega_o \) representing the damping rate and resonance frequency of the corresponding mode, respectively. The cooling efficiency of the CM motion is constrained by the quality factor of the bosonic modes in the system, defined as \( Q_o = \omega_o / \gamma_o \). Although theoretical predictions suggest that the quality factor \( Q_c \) for a magnetically levitated YIG sphere can exceed \( 10^{12} \) under ultra-high vacuum conditions (e.g., \( P = \SI{0.5e-10}{\milli\bar} \))~\cite{2013Thermal,12}, experimentally measured values, which typically range from \( 10^3 \) to \( 10^7 \) in micron-scale
micromagnet spheres, fall significantly short of this theoretical limit. 
The dynamics of this hybrid quantum system, taking into account the dissipation and noise effects, can be described by the following quantum Langevin equations (QLEs):
\begin{align}
	\dot{\hat{a}} = &-\Bigl(\frac{\gamma_a}{2} + i\Delta_a\Bigr)\hat{a} + i g_\text{amc} \sin(kx_0) \hat{m}(\hat{c} + \hat{c}^\dagger)  \nonumber \\
	& - i g_{am} \cos(kx_0) \hat{m} +\Omega_a - 2i \varepsilon_a \hat{a}^\dagger +\sqrt{\gamma_a} \hat{a}^{\text{in}}, \nonumber \\
	\dot{\hat{m}} = &-\left(\frac{\gamma_m}{2} + i\Delta_m\right)\hat{m} + i g_\text{amc} \sin(kx_0) \hat{a}(\hat{c} + \hat{c}^\dagger) \nonumber \\
	& - i g_{am} \cos(kx_0) \hat{a} +\varepsilon_m +\sqrt{\gamma_m} \hat{m}^{\text{in}}, \nonumber \\
	\dot{\hat{c}} = &-\left(\frac{\gamma_c}{2} + i\omega_c\right)\hat{c} + i g_\text{amc} \sin(kx_0)\left(\hat{a}\hat{m}^\dagger + \hat{a}^\dagger\hat{m}\right) \nonumber \\
	& +\sqrt{\gamma_c} \hat{c}^{\text{in}}, \label{QLE}
\end{align}
where \( \gamma_a \), \( \gamma_m \), and \( \gamma_c \) are the decay rates of the cavity, magnon, and CM modes, respectively; \( \hat{a}^{\text{in}} \), \( \hat{m}^{\text{in}} \), and \( \hat{c}^{\text{in}} \) denote their corresponding input noise operators.
Due to the external squeezed vacuum field that drives the cavity mode, extracavity squeezing noise is generated. Consequently, the cavity field effectively evolves within a squeezed environment~\cite{PhysRevA.94.051801,PhysRevA.110.063520}. The quantum noise correlations for the cavity mode satisfy:
$
\langle \hat{a}^{\text{in}\dagger}(\omega) \hat{a}^{\text{in}}(\omega') \rangle = \delta(\omega - \omega') n_s, $
$	\langle \hat{a}^{\text{in}}(\omega) \hat{a}^{\text{in}}(\omega') \rangle = \delta(\omega + \omega') m_s e^{-2i\phi_s},
$with \( n_s = \sinh^2(r_s) \), \( m_s = \cosh(r_s)\sinh(r_s) = \sqrt{n_s(n_s+1)} \), where \( r_s \) is the squeezing parameter and \( \phi_s \) is the squeezing phase. The noise correlation functions for the magnon and CM modes are given by:
$
\langle \hat{m}^{\mathrm{in}}(t) \hat{m}^{\mathrm{in}\dagger}(t') \rangle = (\bar{n}_m + 1) \delta(t - t'), $$ \langle \hat{m}^{\mathrm{in}\dagger}(t) \hat{m}^{\mathrm{in}}(t') \rangle = \bar{n}_m \delta(t - t'),$
$	\langle \hat{c}^{\mathrm{in}}(t) \hat{c}^{\mathrm{in}\dagger}(t') \rangle = (\bar{n}_c + 1) \delta(t - t'), $$\langle \hat{c}^{\mathrm{in}\dagger}(t) \hat{c}^{\mathrm{in}}(t') \rangle = \bar{n}_c \delta(t - t').
$
Due to the nonlinear interaction terms in Eq.~\eqref{QLE}, obtaining direct solutions is challenging.~Therefore, we employ a linearization approach, decomposing the operators into their steady-state average values and quantum fluctuations: \( \hat{o} = o_0 + \delta\hat{o} \) (\( \hat{o} = \hat{a}, \hat{m}, \hat{c} \)). Neglecting second-order fluctuation terms, we obtain the steady-state equations:
\begin{align}
	0 = &-\left( \frac{\gamma_a}{2} + i\Delta_a \right)a_0 - i\left( G_{am} + 2G_{\mathrm{amc}} \operatorname{Re}[c_0] \right)m_0 \nonumber \\
	& + \Omega_a - 2i\varepsilon_a a_0^*, \nonumber\\
	0 = &-\left( \frac{\gamma_m}{2} + i\Delta_m \right)m_0 - i\left( G_{am} + 2G_{\mathrm{amc}} \operatorname{Re}[c_0] \right)a_0 + \varepsilon_m, \nonumber \\
	0 = &-\left( \frac{\gamma_c}{2} + i\omega_c \right)c_0 - iG_{\mathrm{amc}} \left( a_0 m_0^* + a_0^* m_0 \right),
	\label{eq:steady_state}
\end{align}
where the coupling strengths \( G_{am} = g_{am}\cos(kx_0) \) and \( G_{amc} = -g_{amc}\sin(kx_0) \) depend on the equilibrium position \( x_0 \) of the YIG sphere. Recent advances in magnet levitation demonstrate that precise control of the equilibrium position \( x_0 \) can be achieved. When a YIG sphere is positioned at the antinode of the cavity magnetic field, \( k x_0 = (2n+1)\pi/2 \)\hspace{0.5em}, the magnetic field gradient is maximized~\cite{PhysRevLett.123.107702,PhysRevLett.113.156401,PhysRevLett113083603}. The magnon-photon coupling is suppressed (\( G_{am} = 0 \)), while the magnon-photon-CM coupling reaches its maximum value, yielding \( G_{\mathrm{amc}} = -g_{\mathrm{amc}} \).
The linearized QLEs governing the dynamics of the fluctuation operators will be analyzed in the next section.
\section{The effective hamiltonians} \label{4}


The cooling mechanism within our hybrid system can be dynamically controlled by the relative strength of the external drives. By selectively tuning the amplitudes of the cavity drive (\(|\Omega_a|\)) and the magnon drive (\(|\varepsilon_m|\)), our scheme enables flexible switching between two cooling mechanisms: the MCM single-channel  mechanism and the CMI dual-channel mechanism.  
We note that both cooling mechanisms operate outside the multistable regime. 
The dynamics of the MCM single-channel are described as follows. 

(i) The MCM single-channel cooling mechanism.  When the cavity driving strength is significantly stronger than the magnon driving (\(|\Omega_a| \gg |\varepsilon_m|\)), the steady-state solution satisfies the following equations
$	a_0 = \frac{\Omega_a-2i\varepsilon_a a^*_0 }{\frac{\gamma_a}{2} + i \Delta_a}, $ $ m_0 = c_0 \approx 0.$
The corresponding effective Hamiltonian is
\begin{align}
H_{\text{eff}}/\hbar & = \Delta_a \delta\hat{a}^\dagger\delta\hat{a} + \Delta_m \delta\hat{m}^\dagger\delta\hat{m} + \omega_c \delta\hat{c}^\dagger\delta\hat{c} \nonumber \\
&\quad + J_{mc} (\delta\hat{m}^\dagger + \delta\hat{m})(\delta\hat{c} + \delta\hat{c}^\dagger)  + (\varepsilon_a \delta \hat{a}^{\dagger 2} + \varepsilon_a^*\delta \hat{a}^2 ).
\label{eq:magnon_Hamiltonian}
\end{align}
It is dominated by the direct magnon-CM (MCM) coupling term,  $J_{mc} (\delta\hat{m}^\dagger + \delta\hat{m})(\delta\hat{c} + \delta\hat{c}^\dagger)$, where
\(J_{mc} = G_{\text{amc}}|a_0|\).
In this mechanism, cooling is achieved primarily through energy transfer via the MCM interaction. We therefore identify this architecture as the MCM single-channel cooling mechanism. 

(ii) The CMI dual-channel cooling mechanism.  Under the condition \( |\Omega_a| \approx |\varepsilon_m| \),
the steady-state solution satisfies the following equations:
\begin{align}
a_0 &= \frac{-\Omega_a + i\left[ 2\varepsilon_a a^*_0 +  2G_{\mathrm{amc}}\operatorname{Re}[c_0]m_0\right]}{\Delta'_a}, \nonumber \\
m_0 &= \frac{  2iG_{\mathrm{amc}} \operatorname{Re}[c_0] a_0 - \varepsilon_m }{ \Delta'_m }, \nonumber \\
c_0 &= \frac{iG_{\text{amc}} (a_0 m_0^* + a_0^* m_0)}{\omega'_c}, \label{eq:steady_solutions_full}
\end{align}where \( \Delta'_a = -\left(\frac{\gamma_a}{2} + i\Delta_a\right) \), \( \Delta'_m = -\left(\frac{\gamma_m}{2} + i\Delta_m\right) \), and \( \omega'_c = -\left(\frac{\gamma_c}{2} + i\omega_c\right) \).
The effective linearized Hamiltonian
\begin{align}
	\hat{H}_{\mathrm{lin}} = &\Delta_a \delta \hat{a}^\dagger \delta \hat{a} + \Delta_m \delta \hat{m}^\dagger \delta \hat{m} + \omega_c \delta \hat{c}^\dagger \delta \hat{c} \nonumber \\
	& + J_{ac} (\delta \hat{a}^\dagger + \delta \hat{a}) (\delta \hat{c}^\dagger + \delta \hat{c}) \nonumber \\
	& + J_{mc} (\delta \hat{m}^\dagger + \delta \hat{m}) (\delta \hat{c}^\dagger + \delta \hat{c}) \nonumber \\
	& + J_{am} (\delta \hat{a}^\dagger \delta \hat{m} + \delta \hat{a} \delta \hat{m}^\dagger) + \varepsilon_a \delta \hat{a}^{\dagger 2} + \varepsilon_a^*\delta \hat{a}^2,
	\label{eqfull}
\end{align}
where \(  J_{ac}=G_\text{amc} |m_0| \) is the cavity-CM (CCM) coupling strength, \(  J_{mc}=G_\text{amc} |a_0| \)  is the MCM coupling strength, and \( J_{am}= 2G_\text{amc}\operatorname{Re}[c_0]  \) is the cavity-magnon coupling strength.
This effective Hamiltonian  describes a dual-channel cooling mechanism: the interference between the CCM and MCM channels, known as the CMI mechanism. In this process, both the CCM and MCM channels are simultaneously utilized to extract energy from the CM mode.
Crucially, the quantum interference between these two channels destructively suppresses the Stokes (heating) scattering while constructively enhancing the anti-Stokes (cooling) scattering. This targeted manipulation of quantum pathways results in a significant improvement in the net cooling efficiency, which is the hallmark of the CMI mechanism.

\section{Noise spectral density and net cooling
rates}\label{V}
In this section, we analyze the cooling dynamics of the CM mode by deriving the steady-state average occupancy, the quantum noise spectral density, and the net cooling rate. Within the weak-coupling approximation, the steady-state average occupancy of the CM mode $\hat{c}$ is ~\cite{PhysRevLett99093901,Phys.Rev.A.77033804}
\begin{align}
n_{c} \approx \frac{\gamma_{c} \bar{n}_{c} + \Gamma_{+}}{\gamma_{c} + \Gamma_{\text{c}}},
\label{eq:steady_state_population}
\end{align}
where $\Gamma_{\text{c}} \equiv \Gamma_{-} - \Gamma_{+}$ represents the net cooling rate of the CM mode. The coefficient $\Gamma_{-} = S_\mathrm{F_o}(\omega)$ characterizes the cooling rate dominated by anti-Stokes scattering processes, while $\Gamma_{+} =  S_\mathrm{F_o}(-\omega)$ corresponds to the heating rate governed by Stokes scattering processes.
Here, 
$S_\mathrm{F_o}(\omega)$ is the power spectral density (PSD) function that quantifies the quantum noise characteristics of the bosonic mode $\hat{o}$. Physically, $S_\mathrm{F_o}(\omega)$ is the PSD of fluctuations in the quadrature operator $F_o = -G(\delta \hat{o}^\dagger + \delta \hat{o})$ at frequency $\omega$ ($G$ denotes the effective coupling strength between the subsystems), defined by the autocorrelation function
\begin{align}
S_\mathrm{F_o}(\omega) = \int_{-\infty}^{\infty} \mathrm{d}t \, e^{i \omega t} \langle F_o(t) F_o(0) \rangle.
\label{eq:power_spectral_density}
\end{align}
For the hybrid system under consideration, we specifically examine the cavity mode and the magnon mode. The corresponding quadrature fluctuation operators are given by $F_a$ for the photonic subsystem and $F_m$ for the magnon subsystem.
To analyze the cooling performance of the CM mode, we derive analytical expressions for the PSD, the cooling rate, the heating rate, and the net cooling rate of each mechanism.  See \textbf{Supplement 1} for a detailed
derivation of the PSD in two cooling mechanism.

\begin{figure*}
	\centering
	\includegraphics[width=1\textwidth]{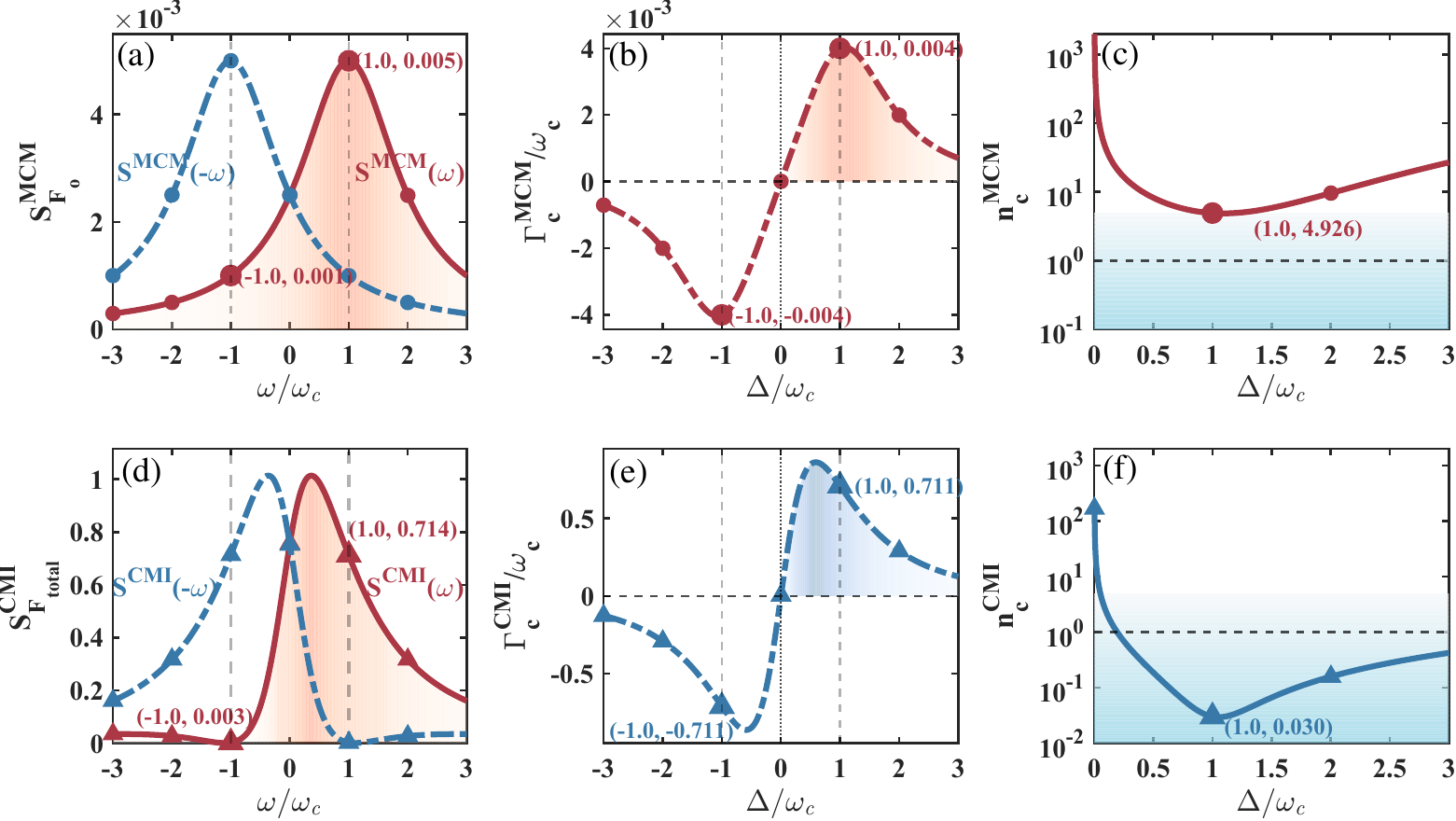}
	\caption{Spectral characteristics and cooling performance comparison. (a, d) The PSD $S_{F_o}(\omega)$ versus the normalized frequency $\omega/\omega_c$.  (b, e) Net cooling rates $\Gamma_c$ and (c, f) $n_c$ as a function of the normalized detuning $\Delta/\omega_c$ (where $\Delta \equiv \Delta_a = \Delta_m$). Panels (a--c) correspond to the MCM mechanism, while (d--f) correspond to the CMI mechanism. Note the orders-of-magnitude enhancement in the cooling rate scale in (e). System parameters: $\omega_c/2\pi = 50$ kHz,   $\gamma_a = 8\omega_c/3$, $\gamma_m = 2\omega_c$, $\bar{n}_m = 0.31$, $J_{ac} = 0.09\omega_c$, $J_{mc} = 0.05\omega_c$, $J_{am} = 0.03\omega_c$, $\gamma_c = 10^{-7}\omega_c$, $r_s = 2$, $\phi_s = 94^{\circ}$, and $\varepsilon_a \approx (0.575 - 0.142i)\omega_c$.}
	\label{fig:SF-wc}
\end{figure*}
\subsection{MCM single-channel cooling mechanism}
\label{subsec:mc}

For the MCM single-channel cooling mechanism, governed by the effective Hamiltonian in Eq.~\eqref{eq:magnon_Hamiltonian}.
The frequency $\omega = \omega_c$ corresponds to the anti-Stokes scattering process, which involves the simultaneous emission of the CM mode and absorption of a magnon, leading to cooling. Conversely, $\omega = -\omega_c$ corresponds to the Stokes scattering process, where the CM mode is absorbed and a magnon is created, resulting in heating~\cite{PhysRevLett128013602}.
The cooling rate and heating rate in the MCM single-channel cooling mechanism  are respectively defined as
\begin{align}
\Gamma_{-}^{\text{MCM}} &= 	S_\mathrm{F_o}^{\text{MCM}}(\omega)
=\frac{{J_{mc}}^2 \gamma_m}{(\omega - \Delta_m)^2 + (\gamma_m/2)^2}, \nonumber\\
\Gamma_{+}^{\text{MCM}}&= 	S_\mathrm{F_o}^{\text{MCM}}(-\omega)
= \frac{{J_{mc}}^2 \gamma_m}{(\omega + \Delta_m)^2 + (\gamma_m/2)^2}.
\label{eq:mc_heating_cooling}
\end{align}

The net cooling rate is given by
$
\Gamma_{\text{c}}^{\text{MCM}} = \Gamma_{-}^{\text{MCM}} - \Gamma_{+}^{\text{MCM}}. 
$
\subsection{CMI dual-channel cooling mechanism}
\label{subsec:cmd}
The CMI dual-channel cooling mechanism utilizes quantum interference between the CCM and MCM channels.  Using the quantum regression theorem~\cite{PhysRev.129.2342,1998Introduction}, we calculate the corresponding cooling and heating rates as follows:
\begin{align}
	\Gamma_{\pm}^{\text{CMI}} = &S_\mathrm{F_{total}}^{\mathrm{CMI}} (\mp\omega) = \gamma_a\left| T^a_\mathrm{SF_a} (\mp\omega) + T^a_\mathrm{SF_m}(\mp\omega_c) \right|^2 \nonumber \\
	& + \gamma_m\left| T^m_\mathrm{SF_a} (\mp\omega) + T^m_\mathrm{SF_m} (\mp\omega) \right|^2 (2\bar{n}_m+1),
	\label{eq:22}
\end{align}
The scattering amplitude $T_{SF_k}^j$ characterizes the transfer of quantum noise from the dissipation channel $j \in \{a, m\}$ (associated with the input noise operators $\hat{a}^{\mathrm{in}}$ and $\hat{m}^{\mathrm{in}}$) to the mechanical force mediated by the system mode $k \in \{a, m\}$. The CMI dual-channel cooling mechanism exploits quantum interference effects between the CCM and MCM channels. The expression for the scattering amplitude $T_{SF_k}^j$ are
\begin{align}
	T^a_\mathrm{SF_a}(\mp\omega) = &\frac{J_{ac}}{S_0} \left[ \left( D(\mp\omega) + 2i \varepsilon_a^* |D(\omega)|^2 \right) \sinh (r_s) e^{-2i\phi_s} \right. \nonumber \\
	& \left. + \left( D(\mp\omega) - 2i \varepsilon_a |D(\mp\omega)|^2 \right) \cosh (r_s) \right], \nonumber \\
	T^m_\mathrm{SF_a}(\mp\omega) = &\frac{J_{ac}}{S_0} \left( -\sqrt{\gamma_m} i J_{am} D_m(\mp\omega) \left[ D^*(\mp\omega) \right. \right. \nonumber \\
	& \left. \left. + 2i \varepsilon_a |D(\mp\omega)|^2 \right] \right), \nonumber \\
	T^a_{\mathrm{SF}_m}(\mp\omega) = &J_{mc}[C_a \sinh(r_s)e^{-2i\phi_{s}} + C^*_{a} \cosh(r_s)], \nonumber \\
	T^m_{\mathrm{SF}_m}(\mp\omega) = &J_{mc} C_m.
	\label{eq:1}
\end{align}
Here, $S_0=1 - 4|\varepsilon_a|^2 D(\omega) D^*(-\omega)$,
$
\frac{1}{D(\omega)} = \frac{1}{D_a(\omega)} + J_{am}^2 D_m(\omega)
$
denotes the optical susceptibility,
where
$
\frac{1}{D_a(\omega)} = \frac{\gamma_a}{2} -i(\omega-\Delta_a ),$
$	\frac{1}{D_m(\omega)} = \frac{\gamma_m}{2} -i(\omega-\Delta_m ),$
$\frac{1}{D_c(\omega)} = \frac{\gamma_c}{2} -i(\omega-\omega_c )
$
represent the response functions of the cavity field, the magnetic field and the CM annihilation operators, respectively. See \textbf{Supplement 1} for remaining parameters.
Equation~\eqref{eq:22} reveals that the total rate is an incoherent sum of contributions from two independent dissipation channels.
The first term, proportional to $\gamma_a$, represents scattering into the  electromagnetic environment via the cavity dissipation channel, while the second term, proportional to $\gamma_m$, corresponds to dissipation into the thermal bath via the magnon dissipation channel.
Here, the scattering amplitudes $T^{a}_{\mathrm{SF}_a}(\mp\omega)$ and $T^{m}_{\mathrm{SF}_a}(\mp\omega)$ correspond to processes mediated by the CCM channel, whereas $T^{a}_{\mathrm{SF}_m}(\mp\omega)$  and $T^{m}_{\mathrm{SF}_m}(\mp\omega)$ correspond to those mediated  by the MCM channel.
Figure~\ref{fig:1}(b) illustrates the distinct scattering processes within these channels.
The quantum interference between the CCM and MCM channels enables the selective suppression of heating processes while enhancing cooling efficiency, which is the cornerstone of the CMI mechanism.

The net cooling rate is given by
$
\Gamma_{\text{c}}^{\text{CMI}} = \Gamma_{-}^{\text{CMI}} - \Gamma_{+}^{\text{CMI}}.
$
The optimal cooling condition is achieved by independently tailoring the quantum interference for the Stokes and anti-Stokes processes. Destructive interference between the two dissipation channels at the Stokes frequency shift ($-\omega$), i.e.,
$
T^a_\mathrm{SF_a} (-\omega) + T^a_\mathrm{SF_m}(-\omega) \to 0, $
$
T^m_\mathrm{SF_a} (-\omega) + T^m_\mathrm{SF_m} (-\omega) \to 0,
$
leads to the suppression of the heating rate $\Gamma_{+}^{\text{CMI}}$. Conversely, constructive interference at the anti-Stokes frequency shift  ($\omega$),
$T^a_\mathrm{SF_a} (\omega) + T^a_\mathrm{SF_m}(\omega) \to \max,$ 
$	T^m_\mathrm{SF_a} (\omega) + T^m_\mathrm{SF_m} (\omega) \to \max,$
results in the maximization of the cooling rate $\Gamma_{-}^{\text{CMI}}$, the net cooling rate $\Gamma_{c}^{\text{CMI}}$ is thereby optimized.

Considering the typically dominant cavity dissipation channel ($\gamma_a \gg \gamma_m$) in many experimental setups, we focus on optimizing the interference condition:
$
T^a_{\mathrm{SF}_a}(-\omega_c) + T^a_{\mathrm{SF}_m}(-\omega_c) \to0.
$
Our numerical optimization tunes parameters ($\phi_s$, $r_s$, $\varepsilon_a$, coupling strengths) to achieve approximately equal amplitudes with a phase difference near $\pi$ between the two scattering paths, thereby strongly suppressing $\Gamma_{+}^{\mathrm{CMI}}$.

\section{Numerical results}\label{VI}
To quantitatively evaluate the performance of our proposed CMI cooling mechanism, we present a detailed numerical analysis comparing it with the conventional MCM mechanism. We evaluate three key parameters as functions of critical system parameters: the quantum noise PSD, the net cooling rate, and the steady-state CM occupancy $n_c$. The results collectively demonstrate that the quantum interference in the CMI mechanism, through constructive and destructive interference, leads to a significant enhancement in cooling performance.
\begin{figure*}[t]  
	\centering
	\includegraphics[width=0.9\textwidth]{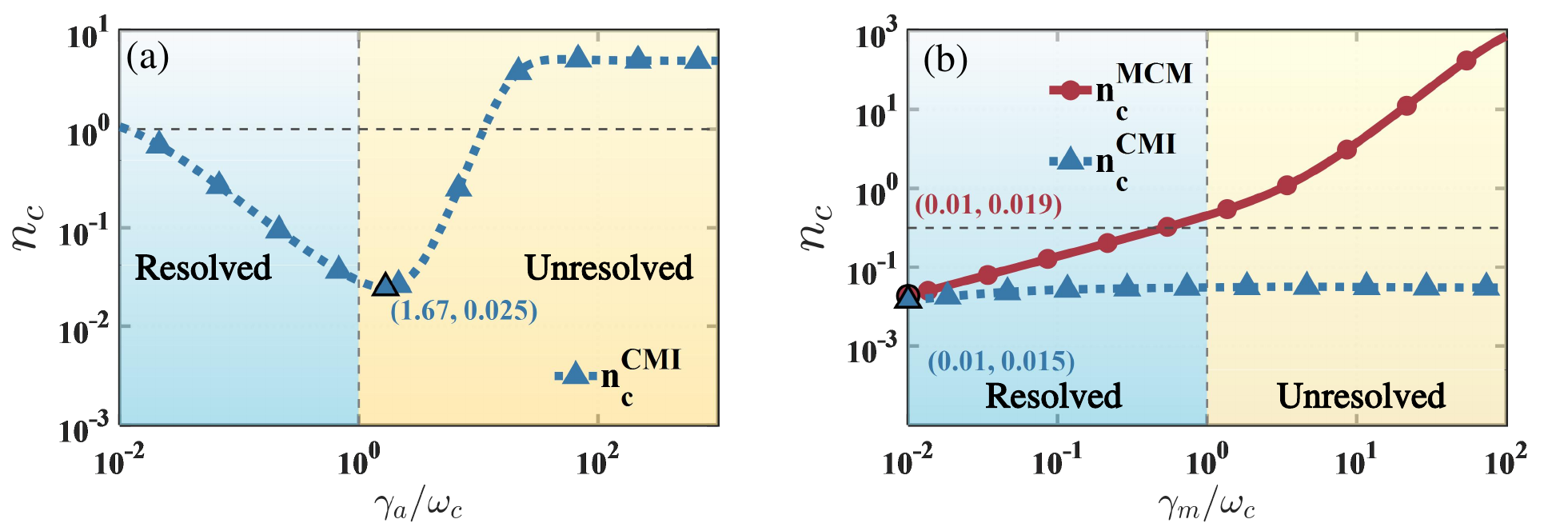} 
	\caption{(a) $n_c$ versus the normalized cavity decay rate $\gamma_a/\omega_c$ for the CMI dual-channel (blue dashed curve) cooling mechanisms. (b)  $n_c$ versus the normalized magnon decay rate $\gamma_m/\omega_c$ for the MCM single-channel (red solid curve) and CMI dual-channel (blue dashed curve) cooling mechanisms.
		The CM quality factor is $Q_c = 10^7$; all other parameters are the same as in Fig.~\ref{fig:SF-wc}.
	}
	\label{nc-r}
\end{figure*}
\subsection{Spectral Properties and CM Cooling Rate}
\label{A}
We begin by examining the quantum noise PSD $S_\mathrm{F_o}(\omega)$, which directly determines the scattering rates. 
Figures~\ref{fig:SF-wc}(a) and (d) display $S_\mathrm{F_o}(\omega)$ as a function of the normalized frequency $\omega/\omega_c$ for the MCM and CMI mechanisms, respectively.
The red and blue curves correspond to the red-sideband
($\Delta_a = \Delta_m = \omega_c$) and blue-sideband ($\Delta_a = \Delta_m = -\omega_c$) conditions.
Under the red-sideband condition, which is a prerequisite for cooling, the PSD exhibits a peak at $\omega = \omega_c$, corresponding to the anti-Stokes (cooling) process, and a dip at $\omega = -\omega_c$, indicating the suppressed Stokes (heating) process.

A quantitative comparison reveals stark contrasts: The MCM single-channel mechanism in Fig.~\ref{fig:SF-wc}(a) exhibits relatively small cooling rate with $S_{\mathrm{F_o}}^{\mathrm{MCM}}(\omega_c) = 0.005$ and heating rate with $S_{\mathrm{F_o}}^{\mathrm{MCM}}(-\omega_c) = 0.001$, respectively. 
In marked contrast, the CMI mechanism in Fig.~\ref{fig:SF-wc}(d) shows a significantly enhanced cooling rate $S_{\mathrm{F_{total}}}^{\mathrm{CMI}}(\omega_c) = 0.714$ and a strongly suppressed heating rate  $S_{\mathrm{F_{total}}}^{\mathrm{CMI}}(-\omega_c) = 0.003$.
This orders-of-magnitude improvement arises from quantum interference between the standard MCM channel and the CCM channel enhanced by an OPA and an injected external squeezed vacuum field.
The distinctive dip at $\omega = -\omega_c$ in Fig.~\ref{fig:SF-wc}(d) serves as a signature of destructive interference, confirming the suppression of Stokes scattering, while the peak at $\omega = \omega_c$ corresponds to constructive interference, verifying the enhancement of anti-Stokes scattering.

The net cooling rate $\Gamma_c$ of the CM mode is plotted as a function of the normalized frequency $\Delta/\omega_c$ in Figs.~\ref{fig:SF-wc}(b) and (e). 
Under the red-sideband condition, the maximum net cooling rate for the CMI mechanism is $\Gamma_c^{\mathrm{CMI}} = 0.711$, which is nearly 180-fold greater than that of the MCM mechanism ($\Gamma_c^{\mathrm{MCM}} = 0.004$).
The cooling rate $\Gamma_-$ exceeds the heating rate $\Gamma_+$, resulting in a net cooling rate $\Gamma_c > 0$ via the sideband cooling mechanism\cite{PhysRevLett.110.153606, PhysRevLett.123.153601,PhysRevLett.117.173602}. 
To quantify the overall cooling performance, we examine the steady-state CM occupancy $n_c$, calculated from Eq.~\eqref{eq:steady_state_population}, and shown in Figs.~\ref{fig:SF-wc}(c) and (f). The minimum steady-state CM occupancies for the MCM and CMI mechanisms are $n_c^{\mathrm{MCM}} = 4.926$ and $n_c^{\mathrm{CMI}} = 0.030$, respectively. This three orders of magnitude reduction in the steady CM occupancy highlights the critical role of quantum interference in the CMI channel, constituting a central result of this work. (We note that all mechanisms exhibit the opposite behavior, i.e., heating, under the blue-sideband condition.)

\subsection{ Minimum steady-state CM occupancy and dissipation rates}
To evaluate the robustness of the cooling mechanisms against environmental dissipation, we analyze the dependence of the steady-state occupancy \( n_c \) on the cavity decay rate \( \gamma_a \) and the magnon decay rate  \( \gamma_m \).

Figure~\ref{nc-r}(a) plots the steady-state CM occupancy $n_c$ versus the normalized cavity decay rate $\gamma_a/\omega_c$.
The CMI dual-channel cooling mechanisms are effective for ground-state cooling ($n_c<1$) across both the resolved-sideband regime ($\gamma_a/\omega_c < 1$) and unresolved-sideband  regime ($\gamma_a/\omega_c> 1$), reaching a minimum occupancy of $n_c^{\mathrm{CMI}} \approx 0.025$ at $\gamma_a/\omega_c = 1.67$.
\begin{figure}
	\centering
	\includegraphics[width=3.3in]{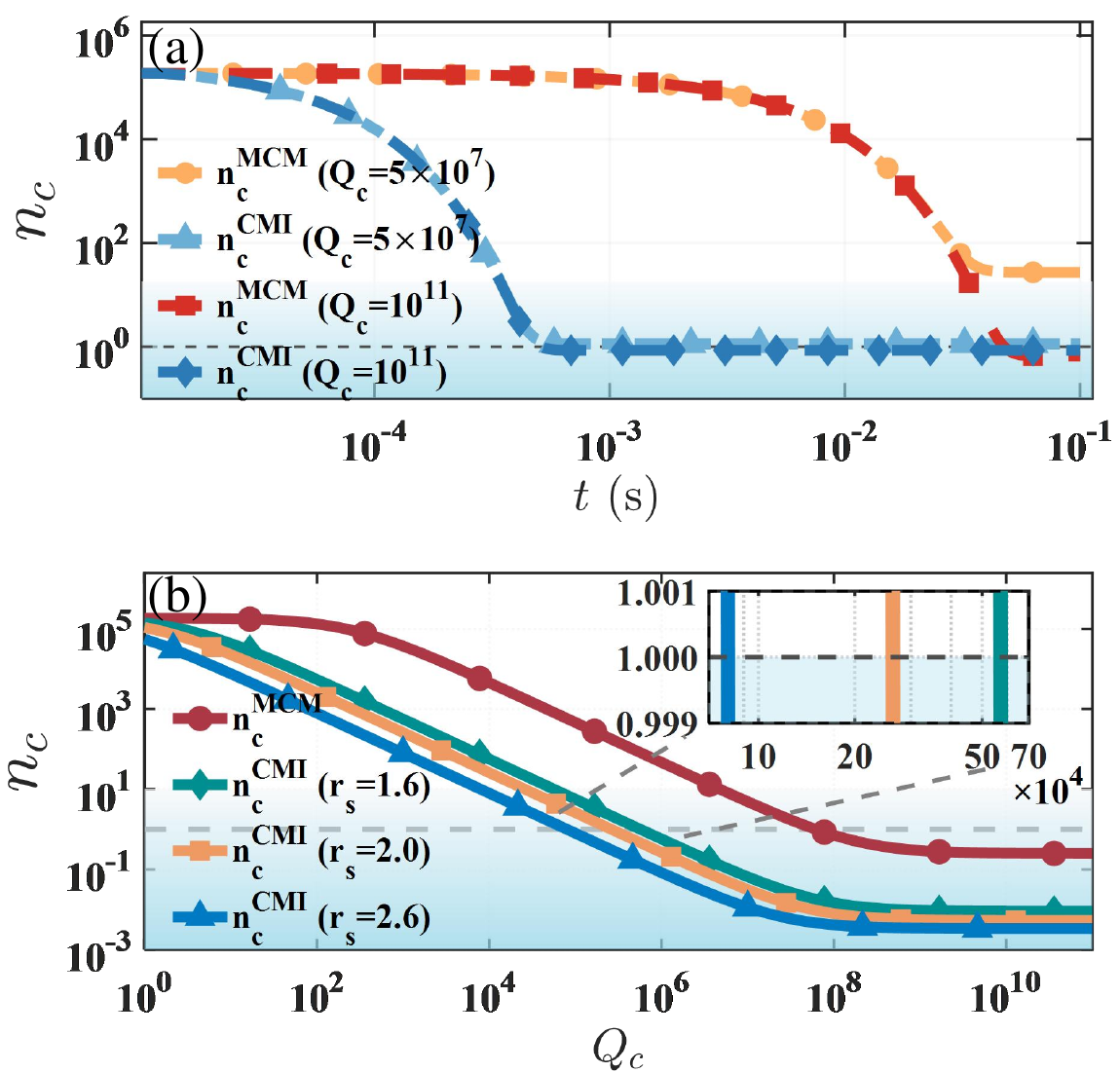}
	\caption{
		(a) Time evolution of the CM occupancy $n_c(t)$ for the MCM and CMI mechanisms. The curves correspond to $Q_c = 5 \times 10^7$ (MCM: yellow circles; CMI: light blue triangles) and $Q_c = 10^{11}$ (MCM: red squares; CMI: blue diamonds). Specific parameters for (a) are $J_{ac} = 0.3\omega_c$, $J_{mc} = 0.025\omega_c$,  and optimal detunings $\Delta_a = \sqrt{\gamma_a^2 + 4\omega_c^2}/2$, $\Delta_m = \sqrt{\gamma_m^2 + 4\omega_c^2}/2$. (b) $n_c$ versus the CM quality factor $Q_c$. The red solid curve denotes the MCM mechanism. The CMI mechanism is shown for squeezing parameters $r_s = 1.6$ (green diamonds), $2.0$ (blue triangles), and $2.6$ (orange squares) with triangular markers. Coupling parameters are $J_{ac} = 0.09\omega_c$, $J_{mc} = 0.05\omega_c$, and $J_{am} = 0.03\omega_c$. The inset highlights the ground-state cooling threshold $n_c \approx 1$. All other parameters are the same as in Fig.~\ref{fig:SF-wc}.
	}
	\label{nc-Qc2}
\end{figure}
We further compare the robustness against magnon dissipation rates $\gamma_m$ for two cooling mechanisms in Fig.~\ref{nc-r}(b).
For the MCM cooling mechanism, $n_c$ remains below unity in the resolved-sideband regime $\gamma_m/\omega_c < 1$ and exceeds 1 in the unresolved-sideband regime $\gamma_m/\omega_c > 1$, confirming the ability to achieve ground-state cooling in the resolved-sideband regime. The minimum average  CM occupancy for this mechanism is $n_c^{\mathrm{MCM}}=0.019$ at $\gamma_m/\omega_c  = 0.01$.
In contrast, the CMI dual-channel mechanism exhibits superior robustness, enabling $n_c < 1$ across both the resolved-sideband regime ($\gamma_m/\omega_c < 1$) and unresolved-sideband regime ($\gamma_m/\omega_c > 1$).

Unlike conventional sideband cooling which relies on filtering out the Stokes scattering via a narrow cavity linewidth (magnon linewidth), our CMI mechanism suppresses the Stokes process via destructive quantum interference. This allows effective cooling even in the unresolved-sideband regime.
This is evidenced by its performance over a broad range of $\gamma_m$, where it maintains a low occupancy, reaching a minimum of $n_c^{\mathrm{CMI}}\approx0.015$ at $\gamma_m/\omega_c  = 0.01$.
This pronounced robustness against both the cavity decay rate \( \gamma_a \) and the magnon decay rate  \( \gamma_m \) makes the CMI mechanism particularly suitable for realistic experimental conditions, where decay rates are often non-optimal.

\subsection{Quality factor analysis}

The quality factor of the CM mode, defined as $Q_c = \omega_c / \gamma_c$, is a critical parameter governing the cooling performance.
A key result of our work is that the CMI mechanism drastically reduces the stringent requirement on $Q_c$.
We first analyze the temporal evolution of the average CM occupancy $n_c(t)$ under different $Q_c$ conditions, as shown in Fig.~\ref{nc-Qc2}(a).
The detailed derivation of \( n_c(t) \) in \textbf{Supplement 1.}
For a high $Q_c$ of $10^{11}$, the MCM cooling mechanism reduces $n_c$ below unity at approximately $8.5 \times 10^{-2}$~s. In contrast, the CMI cooling mechanism achieves the same threshold in merely $8.5 \times 10^{-4}$~s, a two orders of magnitude improvement in cooling speed. Under a more moderate $Q_c$ of $5 \times 10^{7}$, the MCM cooling mechanism fails to reach the quantum ground state ($n_c > 1$), while CMI mechanism remains effective even at $Q_c = 10^{-4}$, still allowing for ground-state cooling within a time scale of $t = 9 \times 10^{-4}\,\text{s}$.\hspace{0.5em}This result clearly demonstrates that the CMI mechanism, by leveraging quantum interference, enables a dramatic increase in energy dissipation efficiency. The resulting rapid cooling is essential for mitigating environmental decoherence, thus facilitating the preparation and coherent manipulation of quantum states in a macroscopic object.

Next, we quantify the experimental feasibility by analyzing the steady-state CM occupancy $n_c$ on
$Q_c$ in two cooling mechanisms.
A detailed analysis of the CMI mechanism, as shown in Fig.~\ref{nc-Qc2} (b).
It reveals that ground-state cooling \(n_c<1\) is maintained over a broad range of \(Q_c\). For a squeezing parameter of $r_s = 2.6$, this range spans from $10^4$ to $10^{11}$. Although the minimum required $Q_c$ increases as $r_s$ decreases (requiring $Q_c > 2.5 \times 10^5$ for $r_s = 2.0$ and $Q_c > 3.6 \times 10^5$ for $r_s = 1.6$), the cooling efficiency remains significant. 
Our numerical simulations employ high squeezing parameters to demonstrate the ultimate potential of the CMI mechanism. However, the results for $r_s=1.6$
in Fig.~\ref{nc-Qc2}(b) confirm that the scheme remains robust and effective at moderate squeezing levels accessible with current technology, despite a reduction in enhancement factors.
In comparison, the conventional MCM mechanism is effective only within a narrow window of 
$10^7 < Q_c < 10^{11}$).	 
A central finding of this work is that the CMI mechanism lowers the minimum required $Q_c$
by more than three orders of magnitude, reducing from the order of $10^7$
to the order of $10^4$.
This drastic reduction significantly relaxes the experimental constraints for preparing macroscopic quantum ground states.

\begin{figure*}[ht]
	\centering
	\begin{minipage}[t]{0.32\textwidth}
		\centering
		\includegraphics[width=\linewidth]{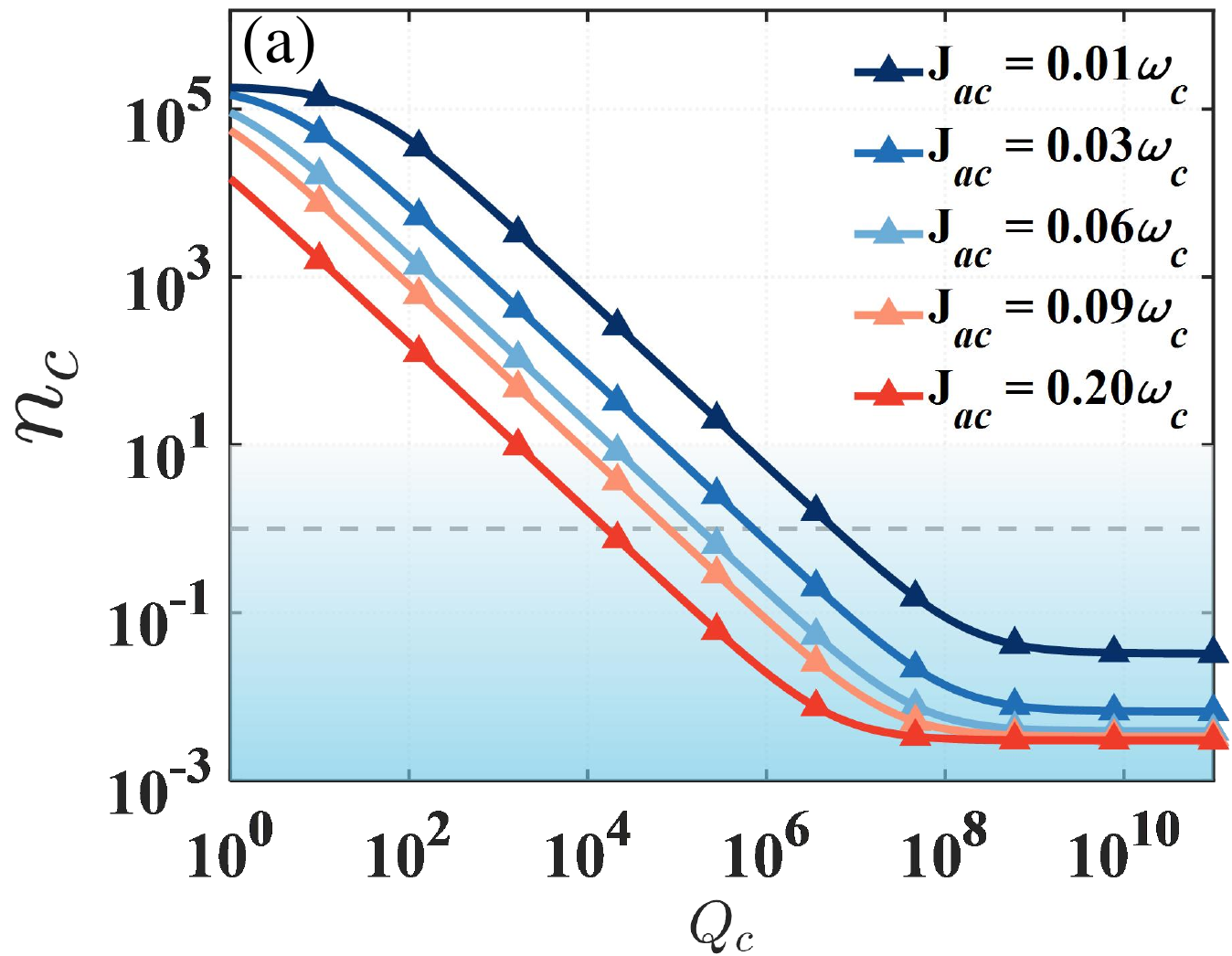}
	\end{minipage}
	\begin{minipage}[t]{0.32\textwidth}
		\centering
		\includegraphics[width=\linewidth]{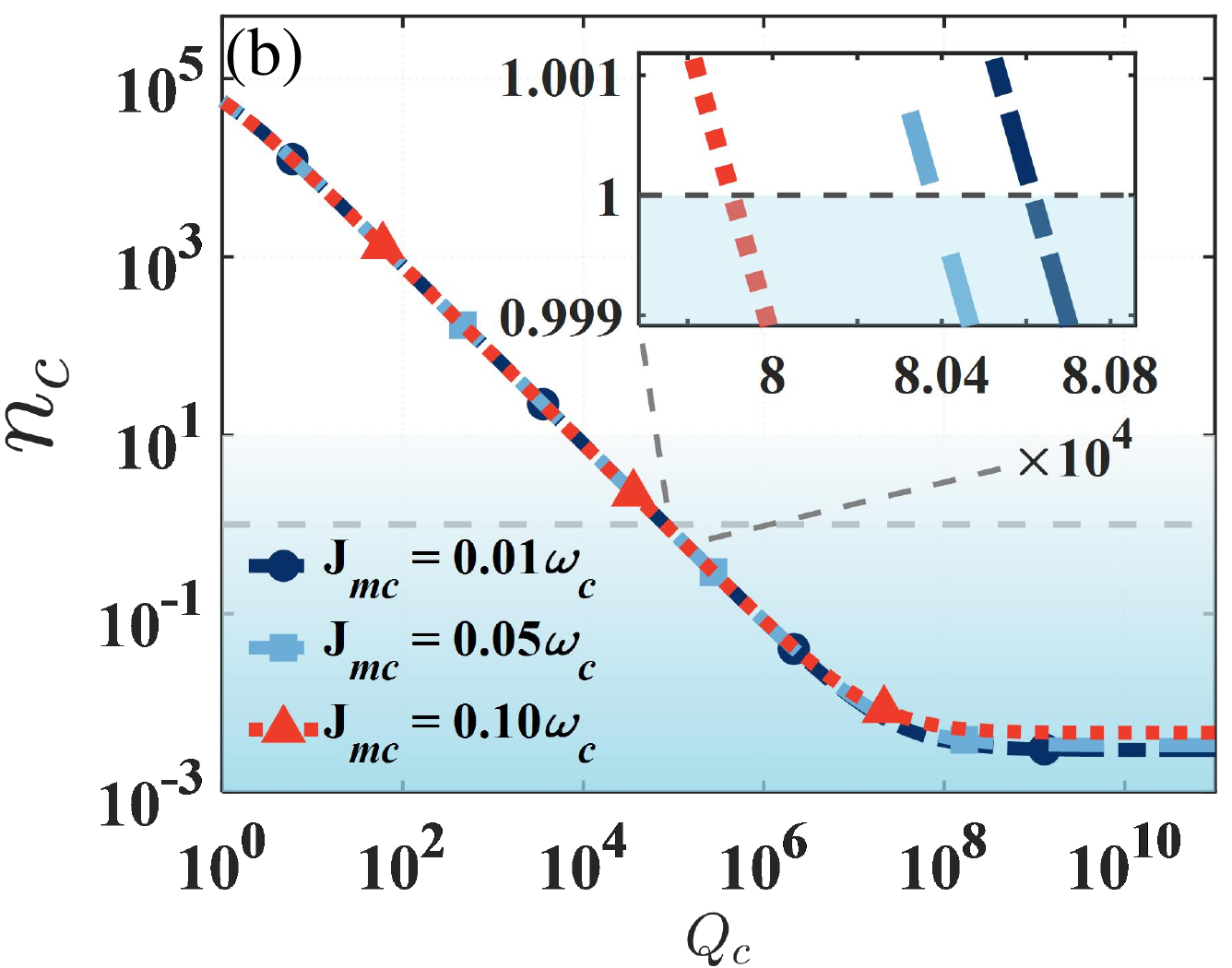}
		\label{fig:2d}
	\end{minipage}
	\hspace{-0.2em} 
	\begin{minipage}[t]{0.32\textwidth}
		\centering
		\includegraphics[width=\linewidth]{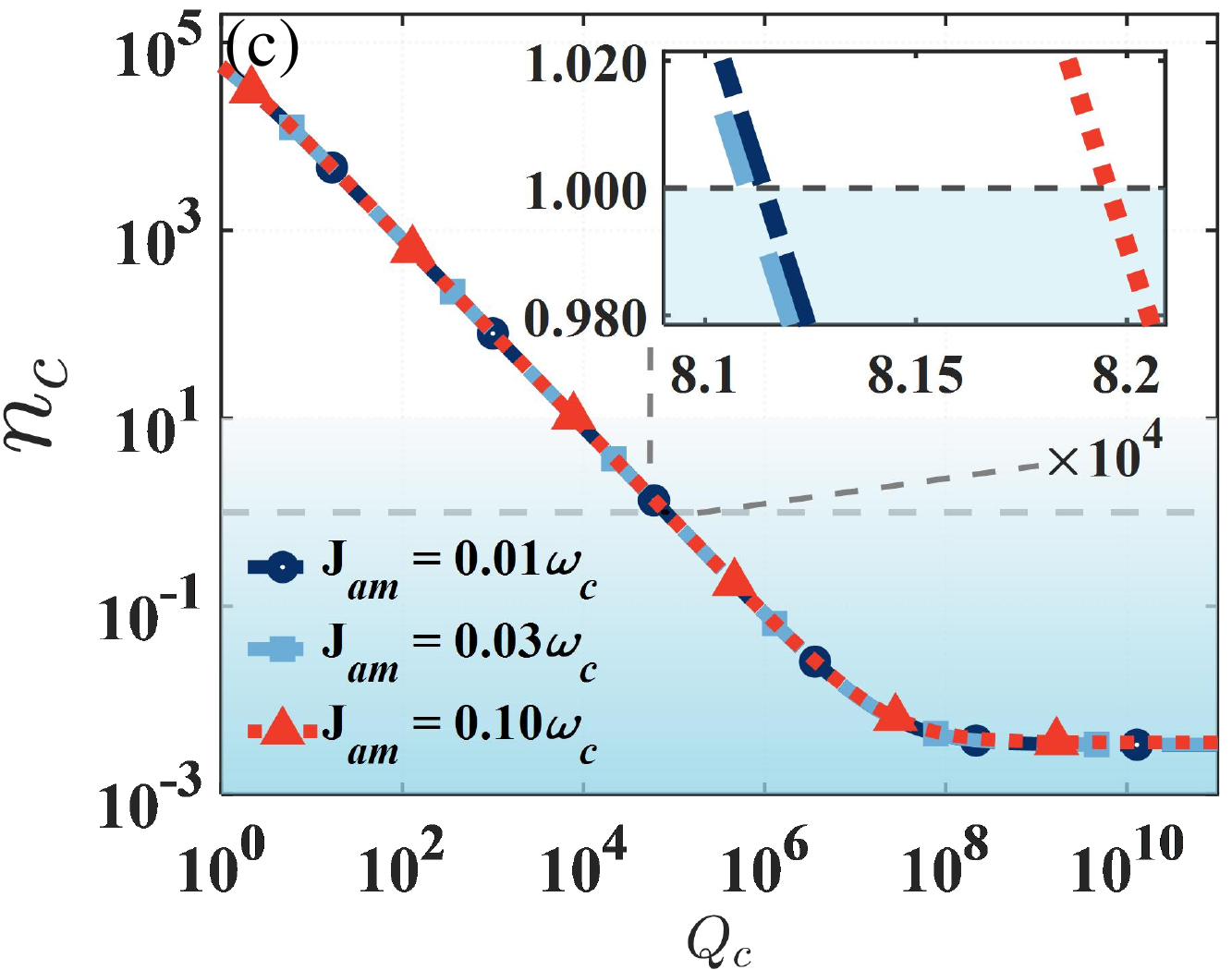}
	\end{minipage}
	
	\caption{
		$n_c$ versus $Q_c$ for different values of (a) the CCM coupling strength $J_{ac}$, with fixed parameters $J_{mc} = 0.09\omega_c$, $J_{am} = 0.05\omega_c$;
		(b) the MCM coupling strength $J_{mc}$,  with $J_{ac} = 0.2\omega_c$ and $J_{am} = 0.05\omega_c$ fixed;
		(c) the cavity-magnon coupling strength $J_{am}$, with the  coupling strengths fixed at $J_{ac} = 0.2\omega_c$ and $J_{mc} = 0.09\omega_c$. The insets in (b) and (c) show zoomed-in views near $n_c\approx1$.
		The squeeze parameter is $r_s=2.6$, compared to Fig.~\ref{fig:SF-wc},
		with other parameters identical to those in Fig.~\ref{fig:SF-wc}.
	}
	\label{nc-Qc}
\end{figure*}

To investigate how coupling strengths affect this $Q_c$ threshold, we analyze its performance under different coupling parameters $J_{ac}$, $J_{mc}$, and $J_{am}$, as summarized in Figs.~\ref{nc-Qc}. These results demonstrate the parametric flexibility of the CMI dual-channel cooling mechanism.

Fig.~\ref{nc-Qc}(a) shows $n_c$ as a function of $Q_c$ for different values of $J_{ac}$. The results indicate that 
the threshold $Q_c$ for ground-state cooling ($n_c < 1$) is highly sensitive to  $J_{ac}$.
As $J_{ac}$
increases from $0.01\omega_c$ to $0.2\omega_c$, the required $Q_c$
drops sharply by over two orders of magnitude. 
Specifically, when $J_{ac}$ is increased from $0.01\omega_c$,  $0.03\omega_c$, $0.06\omega_c$, $0.09\omega_c$, and $0.2\omega_c$, the corresponding threshold $Q_c$ increases markedly from $4.6\times 10^6$ to $6.0\times 10^5$, $7.7\times 10^4$, and $1.7\times 10^4$, respectively.
This underscores that a stronger coupling to the cavity channel can effectively compensate for higher CM dissipation.

In contrast, the dependence on the MCM coupling strength $J_{mc}$
is much weaker, as shown in Fig.~\ref{nc-Qc}(b).
Varying $J_{mc}$ over an order of magnitude results in only a minimal change in the threshold $Q_c$.
Specifically, when $J_{mc}$ is reduced from $0.1\omega_c$ to $0.01\omega_c$, the threshold $Q_c$ rises from $7.99\times 10^4$ to  $8.06\times 10^4$, respectively.
A similarly weak dependence is found for the cavity-magnon coupling strength $J_{am}$ in Fig.~\ref{nc-Qc}(c).
Likewise, a clear decreasing trend of the threshold $Q_c$ with increasing $J_{am}$ is evident. As $J_{am}$ decreases from $0.1\omega_c$ to $0.01\omega_c$, the threshold $Q_c$ increases from $8.11\times 10^4$ to $8.042\times 10^4$ respectively.
These results collectively demonstrate the parametric flexibility of the CMI mechanism. While enhancing any of the coupling strengths lowers the required $Q_c$, the cooling performance can be most effectively optimized by controlling the CCM channel strength $J_{ac}$.
The parametric flexibility, inherent to the CMI mechanism, substantially alleviates the experimental burden of achieving ultra-high $Q_c$. 
\begin{figure*}[ht]
	\centering
	\begin{minipage}[t]{0.325\textwidth}
		\centering
		\includegraphics[width=\linewidth]{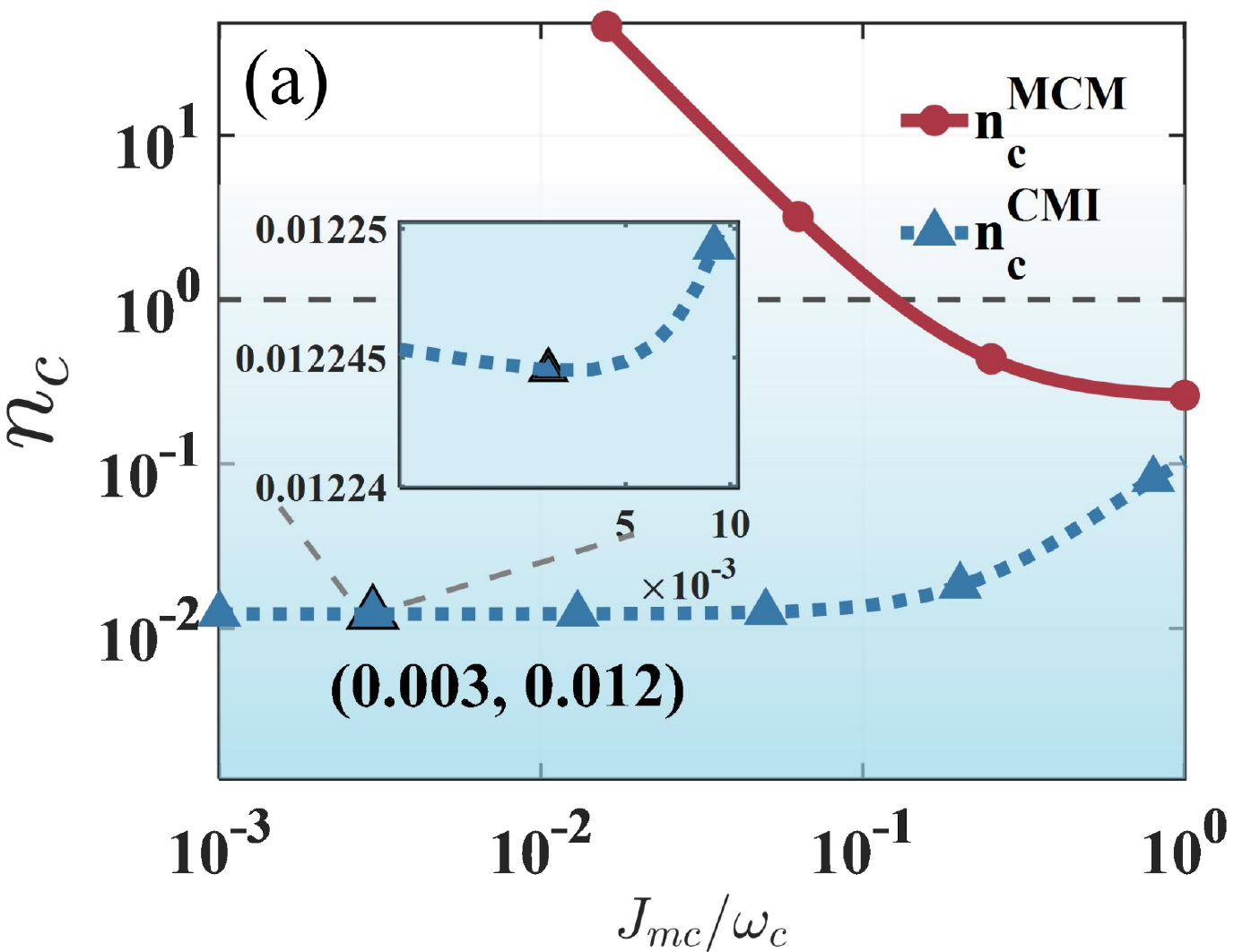} 
	\end{minipage}
	\begin{minipage}[t]{0.325\textwidth}
		\centering
		\includegraphics[width=\linewidth]{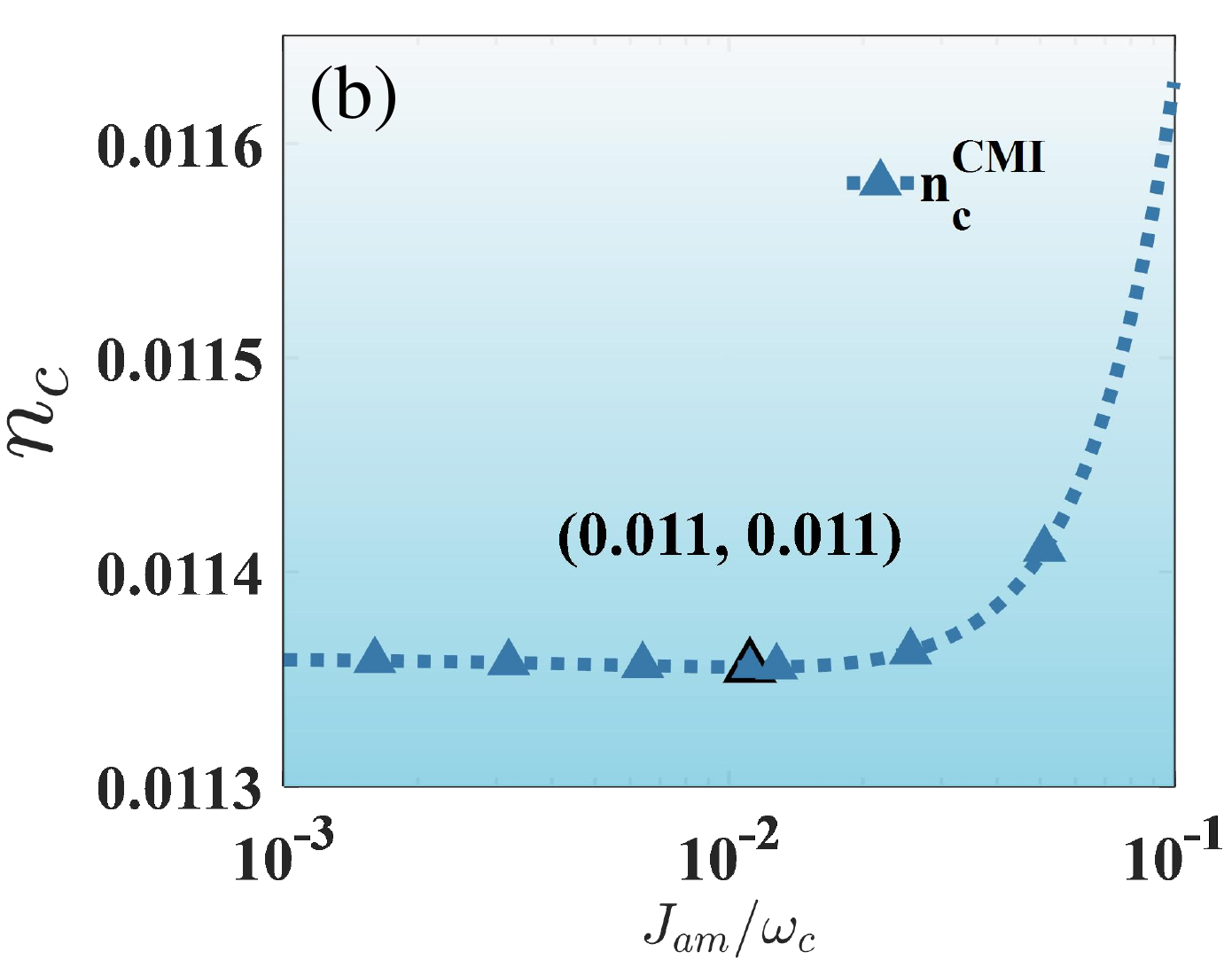} 
	\end{minipage}
	\hspace{-0.2em} 
	\begin{minipage}[t]{0.325\textwidth}
		\centering
		\includegraphics[width=\linewidth]{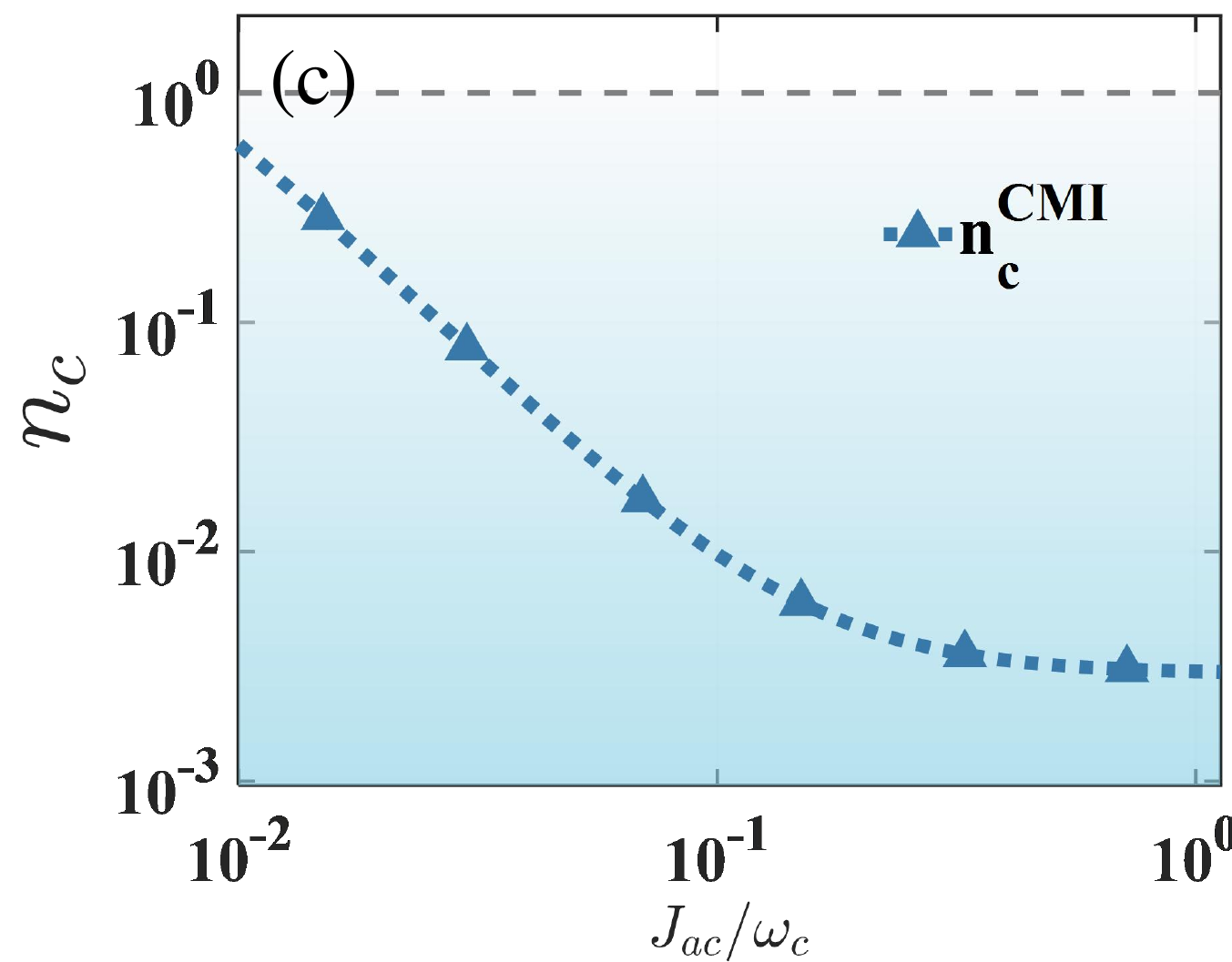} 
	\end{minipage}
	\caption{
		(a)--(c) $n_c$ versus the normalized coupling strengths (a) $J_{mc}/\omega_c$, (b) $J_{am}/\omega_c$, and (c) $J_{ac}/\omega_c$. The insets show zoomed-in views near the cooling threshold. 
		The fixed coupling parameters are:
		for (a) : $J_{ac} = 0.09\omega_c$ and $J_{am} = 0.03\omega_c$;
		for	(b) $J_{ac} = 0.09\omega_c$ and $J_{mc} = 0.05\omega_c$; for (c) $J_{mc} = 0.05\omega_c$, $J_{am} = 0.03\omega_c$.
		In all panels, $\gamma_c = 10^{-5}\omega_c$ and
		the squeeze parameter is $r_s=2.6$.
		Other parameters are the same as in Fig.~\ref{fig:SF-wc}.}
	\label{nc-J}
\end{figure*}

\subsection{Parametric robustness}
\label{subsec:cooling_dynamics}

Furthermore, we examine the robustness of the cooling efficiency against variations in the coupling strengths. Figure~\ref{nc-J}(a) compares the performance of the MCM and CMI mechanisms as a function of the normalized coupling strength
$J_{\text{mc}}/\omega_c$. The conventional MCM mechanism achieves ground-state cooling ($n_c < 1$) within a limited range of $J_{\text{mc}}/\omega_c \in [1.2, 10]$.  In contrast, the CMI mechanism maintains effective cooling over a significantly broader range of $J_{mc}/\omega_c \in [0.001, 1]$, demonstrating superior robustness against variations in this parameter.
We also explore the impact of other coupling parameters specifically on the CMI mechanism. Figure~\ref{nc-J}(b) displays the variation of $n_c$ with $J_{am}/\omega_c$. Ground-state cooling ($n_c < 1$) is maintained across $J_{am}/\omega_c \in [0, 0.1]$. The value of $n_c$ exhibits a non-monotonic dependence, first decreasing to a minimum of $n_c^{\mathrm{CMI}} = 0.011$ at $J_{am}/\omega_c = 0.011$ before increasing again with stronger coupling.  Similarly, Fig.~\ref{nc-J}(c) shows that effective cooling is sustained for $J_{ac}/\omega_c \in [0.01, 1]$. The CMI mechanism's broad operational ranges for all three coupling parameters highlight its exceptional parametric flexibility. This robustness significantly relaxes the experimental constraints, as precise control of coupling strengths becomes less critical for achieving ground-state cooling. 
Additionally, the optimal coupling parameters $J_{ac}$, $J_{mc}$, and $J_{am}$ lie well in the weak-coupling regime. This behavior arises because the steady-state CM occupancy is governed by the net cooling rate, which stems from the competition between cooling and heating processes.

\section{Discussion on experimental implementation}\label{VII}

We now discuss the experimental feasibility of our cooling scheme. 
Our proposal is compatible with state-of-the-art hybrid CMM platforms~\cite{PhysRevLett.113.156401}, where a YIG sphere is coupled to a high-Q superconducting cavity.

The effective coupling strengths used in our simulations are $J_{mc} =G_{\text{amc}}|a_0|= 0.05\omega_c$ and $J_{ac} = G_{\text{amc}}|m_0| = 0.09\omega_c$.
The magnon-photon-CM coupling strength $ G_{\text{amc}}$ (proportional to the field gradient) can be maximized by positioning the YIG sphere at the magnetic field node of the cavity mode, where  the direct magnon-photon coupling $G_{am}$ vanishes, as demonstrated
by Hou and Liu~\cite{PhysRevLett.123.107702}. The effective couplings $J_{mc}$ and $J_{ac}$ are dynamically tunable through the steady-state amplitudes $|a_0|$ and $|m_0|$, which are controlled by the power of the external cavity and magnon drive fields, respectively.
Additionally, the CCM interaction can be efficiently enhanced by directly driving the magnon mode with a microwave field via a loop antenna~\cite{PhysRevLett.120.057202}, which offers higher pumping efficiency compared to indirect driving through the cavity.

A key requirement of our scheme is the generation of both intracavity and extracavity squeezing. The intracavity squeezing can be experimentally realized by terminating the superconducting cavity with a Superconducting Quantum Interference Device (SQUID) or a superconducting nonlinear asymmetric inductive element~\cite{PhysRevA.111.013709,PhysRevLett.119.023602,PhysRevX.7.041011}. 
By modulating the magnetic flux threading the SQUID at twice the cavity frequency, the required squeezing Hamiltonian is generated. Simultaneously, the external squeezed vacuum field can be generated by a separate Josephson Parametric Amplifier or a Traveling Wave Parametric Amplifier operating at cryogenic temperatures. Squeezing levels exceeding 10 dB are routinely achievable in modern superconducting circuit setups, providing sufficient quantum resources for our protocol.

Regarding the system parameters, superconducting coplanar waveguide resonators can achieve ultra-narrow linewidths down to 7.0 kHz~\cite{PhysRevLett.103.043603}.
The magnon frequency $\omega_m$ is widely tunable via an external bias field (covering MHz to 50 GHz~\cite{PhysRevLett.113.156401}), allowing precise resonance matching with the cavity mode.
Since our scheme operates in the microwave regime, the system is housed in a dilution refrigerator (typically at $10\text{--}20~\text{mK}$). While the strong parametric drive required for the CMI mechanism may introduce residual heating, the high thermal conductivity of the YIG sphere at cryogenic temperatures, combined with effective thermalization via the levitating structure, helps mitigate thermal noise. Our numerical results demonstrate that the CMI mechanism remains robust against residual thermal occupancy. Additionally, other decoherence mechanisms in this system are effectively suppressed (as detail in \textbf{Supplement 1}).

Under these optimal conditions, the dominant limitation remains the finite CM quality factor $Q_c$.
Although theoretical models predict that the CM quality factor $Q_c $
can, in principle, exceed $10^{12}$ under ideal conditions~\cite{12}. 
Due to technical limitations, the experimental values of $Q_c$ of micron-sized microspheres typically range from
$Q_c = 10^3$--$10^7$. For instance, Fuwa \emph{et al.} measured \( Q_c \approx 10^3 \) for a YIG sphere (diameter \( 250~\mathrm{\upmu m}\)) via passive magnetic levitation and confinement in a three-dimensional harmonic potential~\cite{PhysRevA.108.063511}, while Hofer \textit{et al.} reported \( Q_c \sim 10^7 \) for a superconducting microspheres~\cite{PhysRevLett.131.043603}.
A major advantage of our CMI mechanism is that it relaxes the stringent requirement on $Q_c$ by three orders of magnitude.
As demonstrated, ground-state cooling is achievable even with
$Q_c \sim 10^4$. 
This significant reduction in experimental constraints paves a realistic path for observing macroscopic quantum phenomena in levitated magnomechanical systems.

\section{Conclusion}
\label{VIII}
In summary,
we have proposed a tunable dual-channel interference enhanced  cooling scheme for improving the CM cooling of a levitated micromagnet in a hybrid CMM system.
This scheme enables flexible switching between the CMI dual-channel and the MCM single-channel cooling mechanisms by independently controlling the drive strengths of the microwave cavity and the magnon mode.

We have demonstrated that the squeezing-enhanced interference between the MCM and CCM channels can give rise to a synergistic cooling effect, significantly increasing the net cooling rate. Specifically, this quantum interference simultaneously enhances the anti-Stokes process while suppressing Stokes scattering, thereby enabling effective cooling even within the unresolved-sideband regime.

Our numerical simulations demonstrate that the CMI dual-channel cooling mechanism yields a net cooling rate of $\Gamma_{c}^{\mathrm{CMI}} = 0.711$, which is approximately 180 times greater than that of the MCM single-channel cooling mechanism with a net cooling rate of $\Gamma_c^{\mathrm{MCM}} = 0.004$.
Moreover, under this cooling mechanism, the steady-state CM occupancy and cooling time are also reduced by two orders of magnitude, reaching $n_c^{\mathrm{CMI}} = 0.030$ and $8.5 \times 10^{-4}$~s, respectively.
Notably, the CMI dual-channel cooling mechanism substantially relaxes the requirement on the CM quality factor, enabling effective ground-state cooling even at $Q_c \sim 10^4$.
This represents a three orders of magnitude reduction in the required
$Q_c$ compared to conventional MCM cooling mechanism.

Our work not only provides a versatile approach to controllable cooling in hybrid quantum systems, but also establishes a theoretical framework for harnessing quantum interference in multimode interactions. These results offer important insights into the quantum control of mesoscopic mechanical oscillators and suggest promising directions for future experimental research, particularly in quantum sensing and the exploration of macroscopic quantum phenomena.\\\\
\bibliography{ref4}
\end{document}